\documentstyle[prl,twocolumn,aps]{revtex}
\input{epsf}

\newcommand{\bq}{\begin{equation}}
\newcommand{\ee}{\end{equation}}
\newcommand{\fr}[2]{\frac{#1}{#2}}
\newcommand{\eps}{\varepsilon}

\begin{document}
\draft

\title{Decay of Quasi-Particle in a Quantum Dot: the role of
Energy Resolution }

\author{P.G.Silvestrov}

\address{Budker Institute of Nuclear Physics, 630090
Novosibirsk, Russia}

\maketitle

\begin{abstract}

The disintegration of quasiparticle in a quantum dot
due to the electron interaction is considered. It was predicted
recently that above the energy $\eps^{*} = \Delta(g/\ln g)^{1/2}$ each
one particle peak in the spectrum is split into many components
($\Delta$ and $g$ are the one particle level spacing and conductance).
We show that the observed value of $\eps^{*}$ should depend
on the experimental resolution $\delta \eps$.  In the broad region of
variation of $\delta \eps$ the $\ln g$ should be replaced by
$\ln(\Delta/ g\delta \eps)$.
We also give the arguments against the delocalization
transition in the Fock space. 
Most likely the number of satellite peaks grows continuously with
energy, being $\sim 1$ at $\eps \sim \eps^{*}$, but remains finite at
$\eps > \eps^{*}$. The predicted logarithmic distribution of
inter-peak spacings may be used for experimental confirmation of the
below-Golden-Rule decay

\end{abstract}


Decay of single-electron excitations in quantum dots became
now the subject of intensive experimental \cite{SImry,Sivan} and
theoretical \cite{Sivan,Blanter,AGKL,Mirlin,Pichard}
investigations. For a closed quantum dot instead of real decay of
a quasiparticle one should consider the disintegration of one
$\delta$-peak in the single-particle spectral density
$\rho(\eps)$ into a relatively dense bunch of peaks. Each
component of this bunch represents the one particle contribution
into complicated exact eigenstate. Quasiparticle life-time in
large Fermi system is usually associated with decay into
two-particle--one-hole configuration. The corresponding width
may be found using the usual Golden Rule \cite{Sivan}. The
energy $\eps'$ at which this width becomes of the same order of
magnitude with three-particle level spasing gives us the
natural threshold for decay of a quasiparticle. However, it was
shown by the authors of Ref. \cite{AGKL} that due to the
effective interaction with five-particle, seven-particle and so
on excitations (they call all states consisting of $n+1$
particles and $n$ holes $2n+1$-st generation) the actual
threshold for particle disintegration is much lower. The more
detailed investigation of statistics of states constituting the
single-particle excitation below $\eps'$ is the subject of this
letter. The statistical approach to finite interacting Fermi
systems has a long history
\cite{Wong,Bohigas,Zelev,Flamb,Shepelyansky}. However, the main
attention was paid to the investigation of the fully developed
chaos. Here we are interested in the
very beginning of the chaotic behaviour, then quasiparticle may
be coupled with only a few many particle states and spacing
between peaks which we consider is much larger than the level
spacing.

The convenient quantity, which describes the splitting
of a noninteracting quasiparticle peak into many peaks, is the
participation ratio (PR) $P = \sum_i \alpha_i^4$. Here $
\alpha_i^2$ is the relative strength of individual peak in
$\rho(\eps)$ and the sum over bunch of peaks corresponding to
one particle excitation is $\sum \alpha_i^2=1$. Physically the
PR is the inverted effective number of exact many particle
eigenstates constituting the quasiparticle excitation.
From technical point of view, the authors of Ref. \cite{AGKL}
have summed up starting from the small excitation energies the
series of special perturbative contributions leading to
quasiparticle disintegration and then estimated, at which energy
$\eps^{*}$ this series blows up. In terms of PR this procedure
gives
\bq\label{PAGKL}
P= 1 - \fr{\eps^2}{g\Delta^2} p(\eps) \ , \ 
p(\eps) = \sum_{n=0} p_n
\left(\fr{\eps^2}{{g}\Delta^2} \ln g\right)^n
 \ ,
\ee
where $\Delta$ is the averaged single-particle level spacing, $g
\gg 1$ is the dimensionless conductance, $\eps \gg \Delta$ is
the energy of our quasiparticle, and $p_n$ are some numerical
coefficients. Each new term in the sum in Eq. (\ref{PAGKL})
corresponds to taking into account more and more complicated
admixtures to quasiparticle. The first term $(n=0)$
describes the mixing with 2-particles and 1-hole,
second to 3-particles and 2-holes, and so on.

At energies close to $\eps^{*} = \Delta(g/\ln
g)^{1/2}$ all terms of the series in Eq. (\ref{PAGKL}) become of
the same order of magnitude. This means that at $\eps >
\eps^{*}$ the PR can not be close to 1. However, the
concrete way of the quasiparticle disintegration with the
growth of $\eps$ depends on the behaviour of the coefficients
$p_n$. In general, the three possibilities for the asymptotics
of $p_n$ are
\bq\label{as}
a). \ p_n \sim n! \ \ ;
b). \ p_n \sim a^n n^\gamma \ \ ;
c). \ p_n \sim 1/n! \ \ ,
\ee
which corresponds to zero, finite and infinite radius of convergence
of the series in Eq. (\ref{PAGKL}) respectively.  One should naturally
expect very different features of the resumed result in these three
cases. Mapping the problem of quasiparticle lifetime onto that of
particle hopping on the Cayley tree led the authors of Refs.
\cite{AGKL,Mirlin} to the asymptotics (\ref{as}b). However, as we will
show below, the actual asymptotics is close to the Eq. (\ref{as}c).

In general, in order to observe experimentally the splitting described
by Eq. (\ref{PAGKL}) one should be able to resolve all many particle
eigenstates, which means that the experimental errors should be
exponentially small $\delta \eps \sim \Delta
\exp(-2\pi\sqrt{\eps/6\Delta})$ \cite{Bohr}. Therefore, in this letter
we would like to find how the mechanism considered in \cite{AGKL} will
manifest itself for more realistic $\delta \eps$. First of all, even
in order to see the decay of quasiparticle into three-particle
configurations one should have sufficiently good resolution $\delta
\eps \sim \Delta^3/\eps^2$ (any few peaks falling into the segment
$\sim \delta \eps$ are seen as one of joint strength $\sum_{\delta
\eps} \alpha_i^2$).  The most interesting is the result for
$\Delta^5/\eps^4 \ll \delta \eps \ll \Delta^3/\eps^2$. Physically this
means that accuracy is much better than needed to resolve the
three-particle levels, but not enough to see the five-particle
ones. In this case
\bq\label{Peps}
P= 1 - \fr{\eps^2}{g\Delta^2} b(\eps) \ , \
b(\eps)=\sum_{n=0} b_n
\left({\eps}/{\eps_c} 
\right)^{2n}
 \ ,
\ee
where $\eps_c = \Delta\sqrt{g/\ln(\Delta/ g\delta
\eps)} $. The transition from pure single-particle to split
spectrum now takes place at $\eps \sim \eps_c$. In particular if
$\delta \eps \sim \Delta^3/\eps^2$ one has $\eps_c
\sim \eps' = \Delta\sqrt{g}$ in accordance with the Golden Rule
prediction \cite{Sivan,AGKL}. At $\delta\eps
\sim \Delta^5/\eps^4$ the expansion (\ref{PAGKL}) formally is
restored, but the coefficients of this new series $p_n^*$ are
much smaller than those of the Eq.  (\ref{PAGKL}).  For better
accuracy $\delta\eps \ll \Delta^5/\eps^4$ the coefficients
$p_n^*$ become a functions of the resolution $p_n^*
=p_n^*(\delta \eps)$. Only at extremely small $\delta \eps $ one
has $p_n^* (\delta \eps \sim \Delta^{n+1} /\eps^{n}) \approx p_n
$.

As it was shown in Refs. \cite{Blanter,AGKL} the values of the
matrix elements 
(MEs) of two-particle interaction are Gaussian distributed with
the variance: 
\bq\label{a}
\overline{ V^2 } = \Delta^2/g^2 \ \ \ .
\ee
Here the numerical factors $\sim 1$ (see e.g. 
Ref. \cite{AGKL}) are included into the definition of
$g\gg 1$ \cite{ref0}. The estimate (\ref{a}) was
done for the diffusive quantum dot.
However, our approach may be valid for the ballistic dot
also. The only necessary condition is that the MEs of
interaction should be random with the amplitude $|V| \ll
\Delta$.

Consider first the mixing of particle with three-particle
states (two particles and one hole). The density ($dn/d\eps$) of
this statesNo spin of quasiparticle.
 is
\bq\label{b}
\nu_3 = {\eps^2}/{4\Delta^3}
\ \ \ .
\ee
Here one factor $1/2$ comes from the integration over the three
energies at fixed $\eps_{p1}  + \eps_{p2} +
\eps_h = \eps$  and another is added due to the Fermi statistics
of two produced particles. We are interested in energies $\eps\ll
\Delta \sqrt{g}$. Therefore $|V| \nu_{3} \ll 1$, which means
that the majority of one-particle states are almost
nonperturbed. In this case the main contribution to the
PR comes from the very small part of levels,
for which the energy difference between one- and three-particles
excitations $\eps^{(1)} -\eps^{(3)}$ occasionally turns out to 
be of the same order of magnitude with the ME $V$. The
relative fraction of such states is small $\sim |V| \nu_{3}$,
but their PR differs by 100\% from $P=1$. Therefore, the
averaged contribution of such events to $P$ is $\delta P_3 \sim
|V| \nu_{3} \sim \eps^2/g\Delta^2$. The accurate calculation
\cite{Ponomarev} allows to find also the numerical factor
\bq\label{d}
P_3 = 1 -2\pi \overline{|V|} \nu_{3} =
 1 - \sqrt{{\pi}/{2}}{\eps^2}/{g\Delta^2}  \ \ .
\ee
Here due to the Eq. (\ref{a}) $\overline{|V|}=\sqrt{2/\pi}
\Delta/g$. 

Mixing of quasiparticle with higher generations
(five-particle, seven-particle etc.) may be formally taken into
account in the same way
\bq\label{e}
\delta P_{2n+1} =-2\pi | V_{eff}^{(2n+1)} | \nu_{2n+1} 
\ \ ,
\ee
where $\nu_{2n+1} \sim \eps^{2n}/\Delta^{2n+1} $. The only
difference from (\ref{d}) is that now $V_{eff}^{(2n+1)}$ is the
effective ME connecting the first and $2n+1$-th
generations via the $n$-th order of the usual perturbation
theory.  The naive estimate of this effective interaction gives
$V_{eff} \sim (\Delta/g)^n (1/\Delta)^{n-1} $, which 
lead to $ \delta P_{2n+1} \sim -(\eps^2/g\Delta^2)^{n} $.  The
main 
advantage of Ref. \cite{AGKL} was in fact the observation that
the high order corrections to $P$ have additional enhancement
$\sim (\ln g)^{n-1}$ compared to the naive estimate. In order to
demonstrate the origin of this large logarithms,
consider the effective ME connecting 
generations 1 and~5
\bq\label{h}
\overline{ |V_{eff}^{(5)}| } =
\overline{ |\sum_2 \fr{V_{12} V_{23}}{\eps_1 -\eps_2}| } = 
\fr{2\Delta^2}{\pi g^2}
\int_{\Delta/g}^{\Delta} \fr{d\eps}{\Delta\eps} =
\fr{2\Delta}{\pi g^2}\ln g \ . 
\ee
Here we have left in the sum over $\eps_2$ only one level
closest to $\eps_1$ (contribution of the other levels is
$\sim 1/\ln g$ smaller) and then averaged over its
position. Therefore, the upper bound of the integral 
is $|\eps|<\Delta$. More interesting is the
origin of the lower bound. The use of the effective interaction
requires $|V_{12}|,|V_{23}| \ll |\eps_1
-\eps_2| \approx |\eps_3 -\eps_2|$. Otherwise one should
consider the strong mixing of three almost degenerate
states $|1\rangle, |2\rangle, |3\rangle$ (accurate taking into
account such a
3-level interaction leads also to $\sim 1/\ln g$ corrections).
That is why the lower bound in the integral in Eq. (\ref{h})
is $|\eps|>\Delta/g$. 

The ME (\ref{h}) together with $\nu_5 \sim \eps^4/\Delta^5$
allows one to reproduce 
the first nontrivial term of the expansion (\ref{PAGKL}).
Moreover, even if one does not take into account the mixing with
higher generations, the correction described by (\ref{h}) could
blow up the PR (\ref{d}) at $\eps \sim \Delta g^{1/2} (\ln
g)^{-1/4}$. However, one more important feature of $\rho(\eps)$
may be 
demonstrated by the Eq. (\ref{h}). The logarithmic divergence of
the integral in (\ref{h}) shows that the MEs with
very different denominators are equally important for the
PR. Suppose we are not able to resolve peaks
which are closer than some $\delta\eps \ll \Delta/g$. As we
mention before, one can see two peaks of comparable amplitude at
distance $\delta\eps$ only if $\delta\eps \sim V_{eff} \sim
\eps_1 - \eps_3$. This means that the upper bound in the integral
in Eq. (\ref{h}) should be chosen via $V_{eff} \sim \Delta^2
/g^2 \eps_{12} >\delta\eps$ and thus
\bq\label{hh}
\eps_{max} = \fr{\Delta^2}{g^2\delta\eps} \ , \ 
\overline{ |V_{eff}^{(5)}(\delta\eps)| } =
\fr{2}{\pi} \fr{\Delta}{g^2}
\ln\left(\fr{\Delta}{g\delta\eps}\right) \ ,
\ee
in accordance with (\ref{Peps}). In order to illustrate this
result we have shown on the Fig. 1 the density of couples of
peaks as a function of the logarithm of spacing $\lambda$
between them $dn/d\ln (\lambda)$ (at $\eps$ slightly below
$\eps^*$). The mixing with third generation leads to the narrow
(width $\sim 1$) peak at $\lambda \sim \Delta^3/\eps^2 \sim
\Delta/g$.  The contribution from fifth generation at
$\lambda \sim \Delta/g$ is in $\eps^2/g \Delta^2$ times weaker,
but such events are uniformly distributed over the wide region
$\ln(\Delta/g^2) < \ln(\lambda) < \ln(\Delta/g)$.

Generalization of (\ref{h}) for an arbitrary generation
gives  
\begin{eqnarray}\label{i}
\overline{ |V_{eff}^{(2n+1)}| } &=&
\left( \fr{\sqrt{2}\Delta}{\sqrt{\pi}g}\right)^{n}
\int^{\Delta} \prod_{i<n}
\fr{d\eps_i}{\Delta \eps_i} 
=
A \fr{2^{\fr{n}{2}} \Delta(\ln
g)^{n-1}}{\pi^{\fr{n}{2}} g^n}
\ , \nonumber\\ 
&\,& \sum_{i=1}^{k} \ln\left(\fr{g\eps_i}{\Delta}\right) > 0 
\ \ , \ \ 
\sum_{i=k}^{n-1} \ln\left(\fr{g\eps_i}{\Delta}\right) > 0
\ ,
\end{eqnarray}
The upper limit for all integrals here is the same as in
(\ref{h}) $\eps_i <\Delta$. The small values of $\eps_i$ are
restricted due to the requirement that none of the
intermediate states in $V_{eff}$ could be mixed strongly with
initial or final state. One may
found the lower and upper bounds for $|V_{eff}|$ by
considering the simplified version of the logarithmic
inequalities in (\ref{i}):
$\ln (g\eps_i/\Delta) > 0$ for any $i$ for lower bound and
$\sum_{1}^{n-1} \ln (g\eps_i/\Delta) > 0$ for upper bound. For
large $n$ such calculation gives
\bq\label{r4}
1 < A < e^n \ . 
\ee
Thus at least the integral (\ref{i}) could not
contain any $n!$ \cite{ref00}. Eqs.
(\ref{e},\ref{i}) together lead to the Eq. (\ref{PAGKL}). 

For finite accuracy one should take into account only the MEs
exceeding the experimental error $V_{eff} > 
\delta\eps$, which is equivalent to the additional restriction
on the domain of integration 
\bq\label{iii}
\sum_{1}^{n-1} \ln \left({g\eps_i}/{\Delta}\right) < 
\ln \left({\Delta}/{g\delta\eps}\right) \ \ .
\ee
If in addition $\ln(\Delta/g\delta\eps) \ll \ln(g)$, the
integration in Eq. (\ref{i}) may be performed explicitly
\bq\label{ieps}
\overline{ |V_{eff}^{(2n+1)}(\delta \eps)| } =
\fr{1}{n-1} \fr{2^{\fr{n}{2}} \Delta}{\pi^{\fr{n}{2}}
g^n}\left(\ln({\Delta}/{g\delta\eps})\right)^{n-1} \ \ .
\ee
In terms of log-distribution of level spacings
$dn/d\ln(\lambda)$ shown on the Fig. 1 the contribution of
generation $2n+1$ leads to correction $\sim
[\ln(\Delta/g\delta\eps)]^{n-1}$.

Consider now the physical consequences for the spectrum of the
different variants of asymptotic behaviour of the coefficients $p_n \
(b_n)$ shown in Eq. (\ref{as}): \\ 
a). In fact, there is no real danger in divergence of the asymptotic
series. One should simply break the summation at the smallest term
(with the number $n_c\sim \sqrt{\eps^*/\eps}$ or $n_c\sim
\sqrt{\eps_c/\eps}$). The same smallest term gives the
order of magnitude estimate of the rest(nonperturbative) part of the
sum. $p(\eps), b(\eps)$ become completely
nonperturbative at $\eps > \eps^*,\eps_c$. However, there is now
indication that it should be $p(\eps>\eps^*)\gg 1$.  One may even see
no considerable disintegration of quasiparticle peaks at $\eps < 
\Delta \sqrt{g}$. \\ 
b). The series in $\eps^2$ has finite radius
of convergence $R=\eps^{*2}/a \ \ (\eps_c^{2}/a) $ and the 
$\gamma$ is responsible
for the kind of singularity of the resummed result at
$\eps^2=R$ (both $a,\gamma \sim 1$). Close to this point all terms 
of the series become equally
important. It is natural to consider such a behaviour as an indication
of the localization-delocalization transition in the Fock space
\cite{AGKL,Mirlin}. \\ 
c). The series is absolutely convergent. We
consider this as the indication of absence of delocalization
transition.

The estimates of ME (\ref{h},\ref{i},\ref{ieps})
were done for a given tree-type Feynman diagram connecting
given initial and final states. Now we have to estimate
the number of such diagrams. First of all, the density of final
states:
\bq\label{nunu}
\nu_{2n+1} = \fr{\eps^{2n}}{\Delta^{2n+1}}
\fr{1}{(2n)!(n+1)!n!} \ \ .
\ee
Here $(2n)!$ appears after the integration over energies of
final particles(holes), $(n+1)!$ and $n!$ account for the
$n+1$ identical particles and $n$ holes. The number of diagrams
for fixed final state is easy to
estimate for the Schr\"odinger perturbation theory. The examples
of diagrams for the screened Coulomb
interaction $V(x-y) \sim \delta(x-y)$ are shown in Fig. 2 
\cite{nonloc}.
Each individual ME of 
$V(x-y)$ corresponds to decay of one particle into two particles
and one hole, ore one hole into two holes and one particle. In 
order
to find the number of diagrams it is convenient to start from
the final state. At first step there are $(n+1)n^2/2$ ways to
join two of 
$(n+1)$ particles and one of $n$ holes into one particle and
$(n+1)n(n-1)/2$ ways to join one particle and two holes into one
hole. Then the same procedure may be repeated with $n$
particles and $n-1$ holes. The number of diagrams
connecting the same initial and final states found in this
way is
\bq\label{numb}
{2^{-n}}{n!(n+1)!(2n-1)!!} \ .
\ee
The doubling of single-particle peaks
is based on very rare events of almost coincidence of
the small ME and small energy difference. This means that the
probability to find two equally large MEs is small and one
should simply multiply the correction (\ref{e}) by the number of
statistically independent diagrams. However, not all of the
diagrams (\ref{numb}) are statistically independent. First of
all, we have not taken into account 
the Fermi statistics of particles in the intermediate states.
This means, that some of the diagrams should cancel each other.
Second, we have estimated the number of diagrams of
Schr\"odinger perturbation theory. If one goes 
to the Feynman technic, many of the diagrams having
the same MEs and different energy denominators will be joined
into one. For example, for two diagrams of Fig.~2 one
has 
\bq\label{Feynman}
{1}/{\eps_a \eps_{b}} 
 +
{1}/{\eps_{a} \eps_c} = 
{1}/{\eps_{b} \eps_c} \ ,
\ee
because $\eps_b +\eps_c=\eps_a$ (for almost degenerate initial
and final states). 
Here $\eps_{a,b,c}$ are the energy denominators for
corresponding cross section on the figure. Therefore, the Eq.
(\ref{numb}) gives only the upper bound of the
number of independent diagrams. 
Combining together (\ref{nunu},\ref{numb}) and the estimate of
$\overline{ |V_{eff}|}$ one finds
\bq\label{pirduha}
p_n < (const)^n 
\fr{n! (n+1)! (2n-1)!!}{(2n)!(n+1)! n!} 
\sim \fr{(const)^n}{n!} \ \ .
\ee
We see that combinatorics of the diagrams (\ref{numb}) could not
compensate the decrease of phase space and the asymptotics of $p_n$
(as well as $b_n$) is described by the Eq. (\ref{as}c).  Slightly
above $\eps =\eps^* ,\eps_c$ due to the mixing with finite number
($\sim \sqrt{\ln g}$) high order (with $n\sim \ln g$) generations the
PR becomes sufficiently smaller than one. This finite number of
connected generations constitutes the main difference of our result
from what happens on the Cayley tree \cite{AGKL,Mirlin}, where even
the first splitting of the quasiparticle peak into two proceeds
through the interaction with all generations. For higher energies our
perturbative approach formally is not valid. We are able to consider
rigorously only the first splitting of quasiparticle peak into two. In
order to go further one should be able to diagonalize exactly the
three-levels almost degenerate events, then the four-levels and so on.
Mathematically, this means that one has to sum up the series of $\sim
1/\ln (g)$ corrections to $P$. Nevertheless, it is natural to suppose,
that further disintegration of quasiparticle also proceeds through the
interaction with finite number of generations. If so, the number of
peaks constituting one excitation most likely will grow smoothly with
energy (crossover instead of phase transition). The delocalization in
the Fock space will not take place in this scenario (although, it may
be difficult to find the experimental evidence of presence or absence
of such delocalization).

Even more informative than $P$ is the distribution of spacings inside
the quasiparticle bunch. The distribution of spacings for first decay
into two peaks (two distinct bunches) has complicated hierarchical
structure. The natural variable to describe this distribution is $\ln
\lambda$ (Fig.  1). In particular this means that the disintegration
threshold $\eps_c$ should depend on the experimental accuracy. It is
natural to expect that this log-distribution of spacings will survive
after further disintegration into three and more peaks. Moreover,
both new delocalization thresholds $\eps^*$ and $\eps_c$ differ only
by the square root of the logarithm from the Golden Rule prediction,
which makes them quite difficult to be observed in the direct
experiment. However the wide logarithmic distribution of spacings
within the single particle bunch of peaks (like that on the Fig. 1)
may be easily distinguished from e.g. Poisson or Wigner-Dyson
distribution. Thus we may conclude that the investigation of spacings
distribution in the single particle spectral density should open the
easiest way to observe the below-Golden-Rule decay of quasiparticles
in quantum dot predicted in Ref. \cite{AGKL}.  Also, the further
investigation of quasiparticle decay may be performed numerically.

Author is thankful to V.~F.~Dmitriev, V.~B.~Telitsin, D.~V.~Savin,
V.~V.~Sokolov and A.~S.~Yelkhovsky for discussions.

\vspace{-1cm}
\begin{figure}
\epsfxsize=8cm
\epsffile{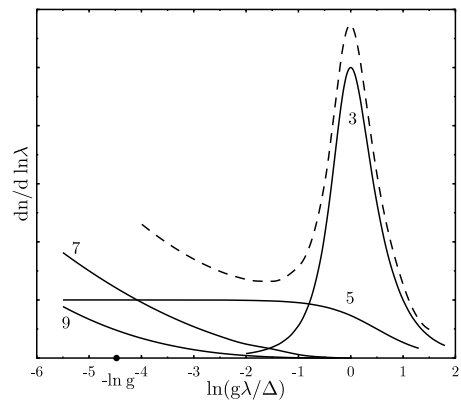}
\vglue 0.2cm
\caption{Distribution of spacings $\lambda$ for first doubling of
the peaks as a function of $\ln \lambda$. The 
mixing with generations 3,5,7,9 is shown.
Dashed line is the total distribution.}
\end{figure}

\begin{figure}
\epsfxsize=6cm
\epsffile{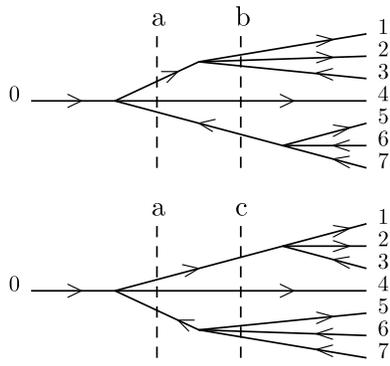}
\vglue 0.2cm
\caption{The examples of diagrams. Energy denominators are
associated with transverse sections (dashed lines).}
\end{figure}


\begin{thebibliography}{99}
\bibitem{ref0} 
Due to \cite{AGKL} for $\eps > E_c \sim \Delta g$ the
variance (\ref{a}) may depend on energy, but we are
interested only in $\eps < \Delta
\sqrt{g}$.
\bibitem{ref00} The authors of
Ref. \cite{AGKL} also instead of estimation of the ME
(\ref{i}) have in fact calculated the upper bound
(\ref{r4}).
\bibitem{nonloc}Taking into account the non-contact interaction
$V(x-y)$ may add the factor $const^n$ to the Eq. (\ref{numb}), but
not $n!$.
\end{thebibliography}
\end{document}